\documentclass{aa}

\usepackage{graphicx}
\usepackage{txfonts}
\usepackage{graphics,graphicx}
\usepackage{amsmath}	% Advanced maths commands
\usepackage{amssymb}	% Extra maths symbols
\usepackage{savesym}
\usepackage{enumitem}
\usepackage{mathrsfs,dsfont}
\usepackage{multirow,multicol}
\usepackage{captcont}
\usepackage{float}
\usepackage[export]{adjustbox}
\usepackage{booktabs}
\usepackage{epstopdf}
\usepackage{soul}
\usepackage{xcolor}
\usepackage[colorlinks=true,linkcolor=blue,citecolor=blue,urlcolor=black]{hyperref}
\usepackage{orcidlink} %this has to come after hyperref
\usepackage{cleveref}
\usepackage{verbatim}
\usepackage{makecell}

\def\be{\begin{equation}}
\def\ee{\end{equation}}
\def\kms{{\rm \,km\,s^{-1}}}

\def\Gyr{{\rm \,Gyr}}

\def\Mpc{{\rm \,Mpc}}
\def\kpc{{\rm \,kpc}}

\def\keV{{\rm \,keV}}

\newcommand{\orcid}[1]{\orcidlink{#1}}

\graphicspath{{figures/}}

\begin{document}

\title{A Minimalist Merger Interpretation of XRISM's Gas Velocity Measurements in the Coma Cluster }

%\subtitle{ }

\author{Congyao Zhang\orcid{0000-0001-5888-7052}\thanks{\email{cyzhang@astro.uchicago.edu}}\inst{\ref{aff1},\ref{aff2}}
\and Eugene Churazov\orcid{0000-0002-0322-884X}\inst{\ref{aff3},\ref{aff4}}
\and Ildar Khabibullin\orcid{0000-0003-3701-5882}\inst{\ref{aff5},\ref{aff4},\ref{aff3}}
\and Natalya Lyskova\inst{\ref{aff4}}
\and\\ Norbert Werner\orcid{0000-0003-0392-0120}\inst{\ref{aff1}}
\and Irina Zhuravleva\orcid{0000-0001-7630-8085}\inst{\ref{aff2}}
%\vspace{-5pt}
}
\institute{
Department of Theoretical Physics and Astrophysics, Masaryk University, Brno 61137, Czechia\label{aff1}
\and Department of Astronomy and Astrophysics, The University of Chicago, Chicago, IL 60637, USA\label{aff2}
\and Max Planck Institute for Astrophysics, Karl-Schwarzschild-Str. 1, D-85741 Garching, Germany\label{aff3}
\and Space Research Institute (IKI), Profsoyuznaya 84/32, Moscow 117997, Russia\label{aff4}
\and Universit\"{a}ts-Sternwarte, Ludwig-Maximilians-Universit\"{a}t M\"{u}nchen, Scheinerstr. 1, 81679 Munich, Germany\label{aff5}
}

\date{Received xxx; accepted xxx
}

\abstract{The recent microcalorimetric X-ray observations of the Coma cluster by XRISM have sparked active discussion regarding the physical origin of its gas velocity features. Here, we demonstrate that an off-axis minor merger in its early phase -- when the infalling subhalo is near its primary apocenter and the stripped tail is not yet mixed with the main cluster atmosphere -- can drive intracluster gas motions generally consistent with the XRISM results. These include a pronounced velocity gradient and an approximately uniform velocity dispersion of $\simeq100-200\kms$ in the cluster core. Our merger scenario was originally suggested in \citet{Lyskova2019} to reproduce the major X-ray morphological features of Coma. In addition, we introduce a simple and robust diagnostic of intracluster gas motions based on the ratio of the line-of-sight velocity to the velocity dispersion.
}


   \maketitle
%
%-------------------------------------------------------------------

\section{Introduction}

The recently-launched XRISM X-ray observatory enables precise velocity mapping of the hot intracluster medium (ICM), primarily thanks to its spatially-resolved high-resolution spectroscopic capabilities of the onboard Resolve microcalorimeter \citep{Ishisaki2022}. The obtained gas-kinematic maps provide new insights into cluster assembly history and baryonic physics \citep[e.g.,][]{XRISM2025_Centaurus,XRISM2025_Perseus,XRISM2025_Coma,Heinrich2025,Zhang2025}.

Among the clusters observed during the first two years of XRISM's operation, the Coma cluster exhibits intriguing velocity features \citep{XRISM2025_Coma,Gatuzz2025}, including (1) a high gas mean velocity in the core ($\simeq-500\kms$) relative to galaxies, (2) a strong velocity gradient, i.e., $\simeq500\kms$ over a $\simeq400\kpc$ spatial scale, and (3) an approximately uniform velocity dispersion ($\simeq200\kms$) across the three observed regions separated by $\simeq200\kpc$ in projection. Implications of these results have been a subject of intense discussions \citep[e.g.,][]{Vazza2025,Eckert2025,Groth2025}. \citet{XRISM2025_Coma} showed that a model of a spatially uniform turbulence with a power law velocity power spectrum requires a much steeper slope compared to the Kolmogorov scaling. However, \citet{Vazza2025} argued that the XRISM measurements are statistically consistent with more realistic power spectra from cosmological simulations and the observed steepening arises from complex “patchy” turbulent structures.

The Coma cluster has long been known as a non-relaxed system due to its rich substructures in both the optical and X-ray wavelengths \citep[e.g.,][]{Fitchett1987,Colless1996,Briel1992,Neumann2001}. Recognizing its merging nature \citep{Burns1994}, we analyze the velocity field predicted by an ongoing minor-merger scenario that we previously developed for Coma in \citet[][see also \citealt{Zhang2019,Sheardown2019}]{Lyskova2019}. Our merger model was primarily constrained by the extended, stripped X-ray tail of the galaxy group NGC~4839 to the southwest, which indicates an on-going off-axis merger between Coma and the group. The latter is supposed to be near its primary apocenter, $\sim1\Gyr$ after the pericentric passage. This picture predicts the presence of a runaway merger shock on the far side of the group and a secondary, trailing shock front, both associated with prominent radio features observed in Coma (e.g., \citealt{Brown2011,Bonafede2022}; see detailed theoretical discussions in \citealt{Zhang2019,Zhang2021}). This scenario was corroborated by the SRG/eROSITA X-ray image of the entire cluster \citep{Churazov2021,Churazov2023}.

\begin{figure*}
\centering
\includegraphics[width=0.95\linewidth]{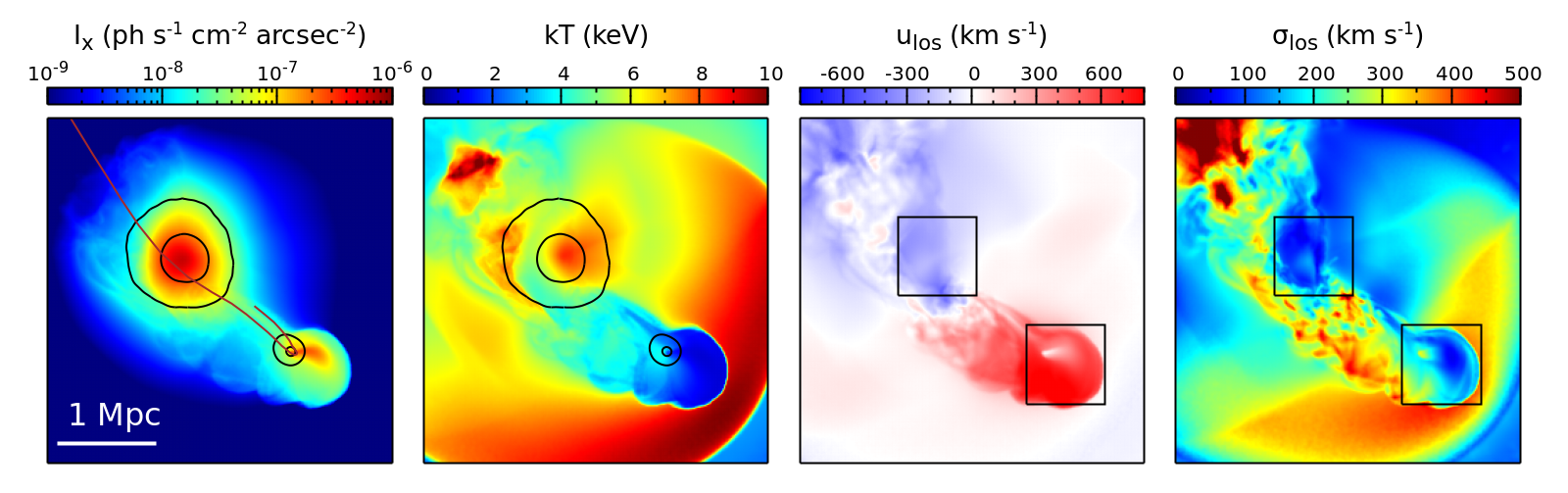}
\vspace{-5pt}
\caption{
An off-axis minor merger between two non–cool-core clusters with a merger mass ratio of $\xi_0=6$. The panels show the simulated X-ray surface brightness ($0.5-8\keV$), X-ray–weighted temperature, LOS bulk velocity, and velocity dispersion at $t=1.0\Gyr$ after the pericentric passage, viewed at an inclination angle of $i\simeq40^\circ$. Black contours in the left panels indicate the total projected mass distribution. The dark-red line marks the trajectory of the subhalo. Zoom-in views of the gas velocity distributions in the regions outlined by the black boxes are shown in Fig.~\ref{fig:merger_proj_zoom}. This simulation well resembles Coma in X-ray observations (see Section~\ref{sec:merger}). }
\label{fig:merger_proj}
\end{figure*}

Revisiting our merger model, we find that this previously-proposed off-axis minor merger successfully reproduces the velocity features observed by XRISM, providing a natural explanation for the large bulk velocity, velocity gradient, as well as the moderate velocity dispersion in the cluster core. In this picture, the observed gas motions are still in its early driven phase at large scales. The turbulent cascade has no time to (fully) develop, as also proposed by \citet{XRISM2025_Coma}. This result underscores the importance of detailed and tailored numerical modeling of individual galaxy clusters to interpret the XRISM velocity measurements. It also offers a simple guideline for selecting best-matching examples from cosmological simulations, including those with constrained initial conditions.

\section{A recent off-axis minor merger in Coma} \label{sec:merger}

Fig.~\ref{fig:merger_proj} shows our merger scenario, resembling the X-ray morphology of Coma. The simulation models a merger between two idealized non-cool-core clusters with the mass ratio of $\xi_0=6$ using the moving-mesh code {\sc Arepo} (\citealt{Springel2010,Weinberger2020}; see Appendix~\ref{sec:appendix:methods} for more details on our numerical setups and experiments of different merger parameters).

In this configuration, the group bypassed the main cluster core from the eastern side and is currently near its primary apocenter, as shown in Fig.~\ref{fig:merger_proj}. The moment of the group's primary pericentric passage is defined as $t=0$. The red line indicates the group's trajectory. Our model reproduces the major X-ray and Sunyaev–Zel’dovich effect \citep{Planck2013} features of the system, including the gaseous tail of the group, the gas bridge connecting the group and the main cluster, as well as multiple identified shock fronts and contact discontinuities (see \citealt{Lyskova2019,Churazov2021} for more detailed comparisons with the X-ray observations).

\begin{figure}
\centering
\includegraphics[width=0.95\linewidth]{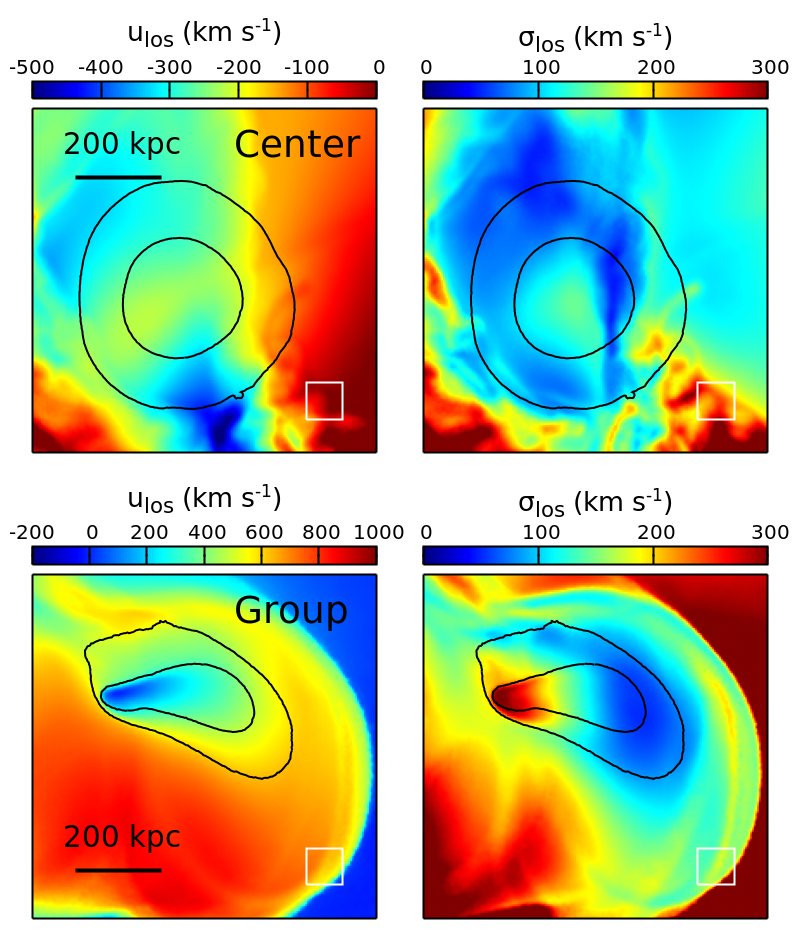}
\vspace{-5pt}
\caption{Zoom-in views of the gas velocity distributions in the core of the main cluster (top) and the infalling group (bottom). Black contours indicate the X-ray surface brightness distribution. The small white box in the bottom-right corner marks the size of the XRISM/Resolve's full field-of-view. The simulation reproduces key velocity features in the core of Coma (see Section~\ref{sec:merger}).  }
\label{fig:merger_proj_zoom}
\end{figure}

The modeled X-ray-weighted velocity distributions along the line-of-sight (LOS) are shown in the right panels, with zoom-in views of the cluster/group cores provided in Fig.~\ref{fig:merger_proj_zoom}. All bulk velocities are measured in the rest frame of the main dark matter halo. Note that we had a relatively loose constraint on the LOS viewing angle based on the morphology of the group's X-ray tail and the velocities of the cluster member galaxies, requiring that the merger plane not deviate significantly from the plane of the sky. XRISM measurements provide additional constraints: the main cluster's gaseous core exhibits a negative velocity (moving towards us) relative to the gravitational potential of the system \citep{XRISM2025_Coma}. We find that an inclination angle of $\sim30^\circ-40^\circ$, defined as the angle between the LOS and the norm of the merger plane, generally resembles the observed velocity structure and major X-ray morphological features mentioned above. The LOS velocity variations with the amplitude of $\sim500\kms$ are visible in the core of the main cluster on scales of a few $100\kpc$, reflecting large-scale bulk velocities induced by the bypassing group. In the same region, the gas velocity dispersion remains below $\sim150\kms$, indicating a lack of small-scale turbulent structures, consistent with the XRISM results. A full turbulent cascade of the merger-driven large-scale motions has not yet developed.
We emphasize that achieving a closer match to the observed velocity structures would require additional fine-tuning of the merger parameters (e.g., mass ratio and impact parameter), the viewing angle, and the merger timing (see Fig.~\ref{fig:merger_proj_params} for an illustration of the parameter dependence). Such fine-tuning is not the focus of this work and would not affect our theoretical picture.

\begin{figure}
\centering
\includegraphics[width=0.95\linewidth]{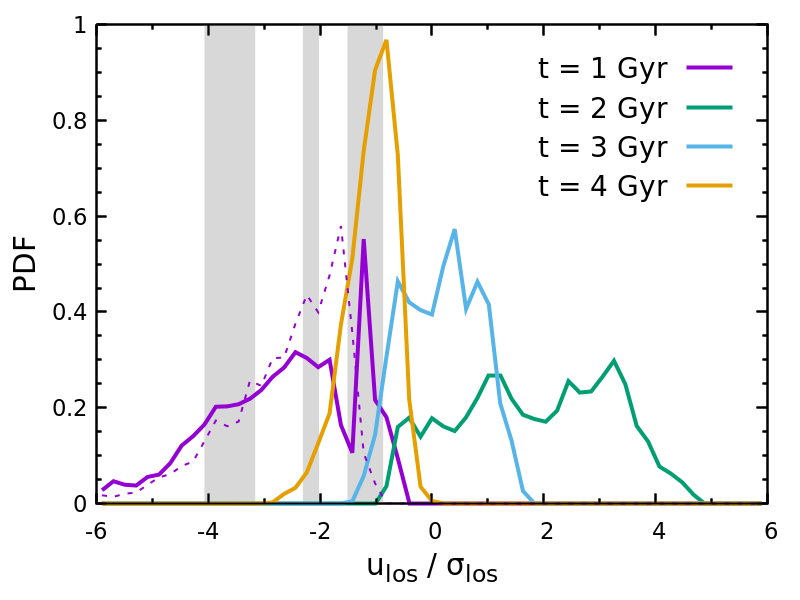}
\vspace{-5pt}
\caption{PDF of the velocity ratio $u_{\rm los}/\sigma_{\rm los}$ obtained in our idealized merger simulation. The solid lines show its time evolution within $r<250\kpc$ of the main cluster. The dashed purple line shows the result from a merger with a larger mass ratio $\xi=10$ (see the bottom panels in Fig.~\ref{fig:merger_proj_params}), indicating that the velocity ratio is relatively insensitive to the absolute velocity amplitude. The shaded grey regions indicate the three XRISM measurements in Coma, supporting our argument that merger-driven, large-scale bulk motions dominate the observed velocity fields in Coma (see Section~\ref{sec:merger}). }
\label{fig:ratio_sim}
\end{figure}

The dimensionless ratio $u_{\rm los}/\sigma_{\rm los}$ provides a robust indicator of gas motion characteristics, which is less sensitive to the absolute velocity amplitude (see Section~\ref{sec:discussion:scale} for more discussions). Fig.~\ref{fig:ratio_sim} shows the time evolution of the probability density function (PDF) of this ratio within $r<250\kpc$ of the main cluster (solid lines). In the early phase, a broad PDF, corresponding to a large bulk velocity gradient, is driven as the infalling group perturbs the velocity field during the core passage. Sloshing-like motions gradually form in the main cluster core, manifested as swinging of the PDF across zero ($t\simeq2-3\Gyr$; see also Appendix~\ref{sec:appendix:sloshing}). As the turbulent cascade gradually develops, the PDF becomes much narrower. For comparison, the three XRISM measurements of this ratio in Coma are shown as shaded vertical regions, ranging from approximately $-4$ to $-1$, which supports our merger interpretation.

At present, the group is moving slowly near the apocenter in the rest frame of the system, as shown in Fig.~\ref{fig:merger_proj}, with a LOS velocity of $\simeq290\kms$ for the dark matter component and slightly slower motion for the gas ($\sim200\kms$). Note that the exact velocity, including its LOS direction, is sensitive to both the group's trajectory and viewing angle. Observationally, the LOS velocity of the NGC~4839 group is $\simeq470\kms$ relative to Coma \citep{Adami2005}. We stress that these apparently subsonic motions do not contradict the presence of a bow shock observed ahead of the group (to its southeast; \citealt{Mirakhor2023}), as the group core is currently being swept by its own stripped tail. The tail is moving rapidly away from us with a LOS velocity of $\sim800\kms$, corresponding to a 3D velocity of $>1000\kms$ relative to the group core, which naturally forms a shock. The relatively high velocity dispersion in the central core of the group ($\simeq250\kms$) is mainly caused by the LOS overlap between the group and its tail. Future XRISM observations may be able to separate these two velocity components, which differ by several $100\kms$ in our simulation.

\section{Discussion}

\subsection{Dimensionless ratio $u_{\rm los}/\sigma_{\rm los}$ as a motion diagnostic} \label{sec:discussion:scale}

\begin{figure}
\centering
\includegraphics[width=1\linewidth]{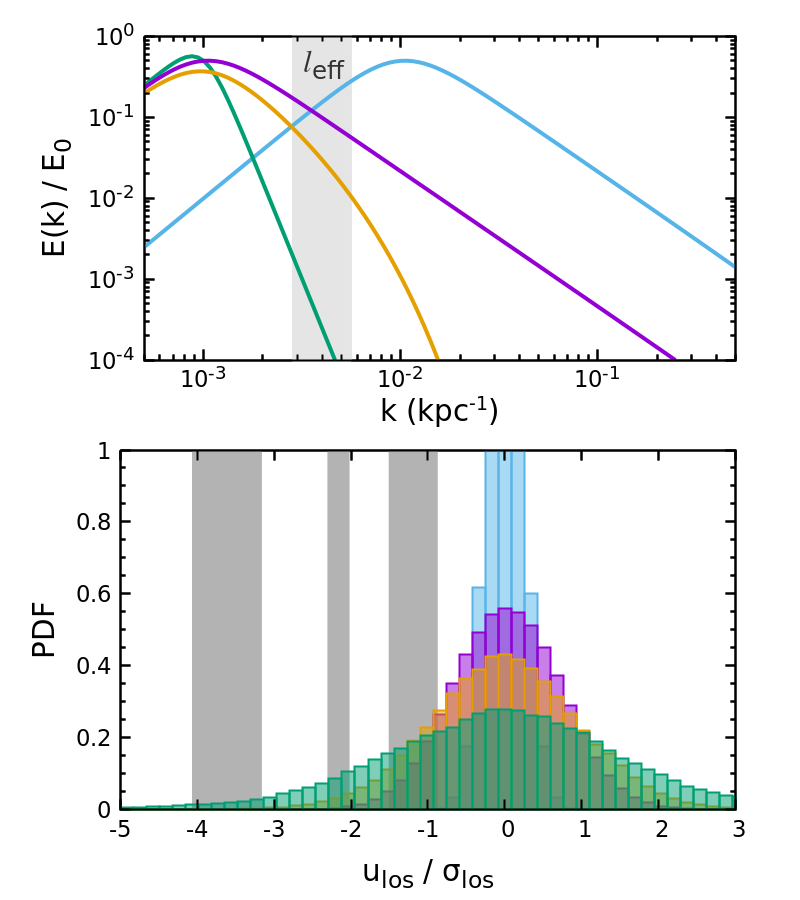}
\vspace{-2pt}
\caption{Velocity-ratio PDFs (bottom) derived from Gaussian random fields with different underlying energy power spectra. The adopted $E(k)$ are illustrated in the top panel with $(\ell_{\rm inj},\,\alpha,\,\ell_{\rm diss})=$ $(1\Mpc,\,-5/3,\,0.1\kpc)$, $(1\Mpc,\,-5/3,\,300\kpc)$, $(1\Mpc,\,-6,\,0.1\kpc)$, and $(0.1\Mpc,\,-5/3,\,0.1\kpc)$ for the purple, yellow, green, and blue lines, respectively. The vertical band in the top panel marks the approximate range of $\ell_{\rm eff}$ in the Coma gas core. The bands in the bottom panel indicate the three XRISM measurements in Coma. These measurements deviate significantly from the predictions, indicating that the cluster is in a merger-driven, transient phase, where velocity distributions depart from Gaussianity (see Section~\ref{sec:discussion:scale}).}
\label{fig:turb_sim}
\end{figure}

The ratio between the gas bulk velocity (relative to the gravitational potential) and the velocity dispersion can serve as a quick and useful diagnostic of the types of gas motions and dominant motion scales in the ICM \citep{Zhuravleva2012}. The idea is to utilize the ``filter effect'' induced by the X-ray emissivity distribution. The majority of the X-ray flux along a given sight-line originates from the LOS portion closest to the cluster core, the spatial scale of which is characterized as the effective length $\ell_{\rm eff}$ \citep{Zhuravleva2012,XRISM2025_Perseus}. Gas motions on scales much smaller than $\ell_{\rm eff}$ contribute primarily to the velocity dispersion, leading to $|u_{\rm los}/\sigma_{\rm los}|\ll1$. Otherwise, we expect to see significant bulk velocities with $|u_{\rm los}/\sigma_{\rm los}|\gtrsim1$. In Coma, the X-ray surface brightness is flat within the core \citep{Churazov2012}, corresponding to an approximately uniform $\ell_{\rm eff}\sim250\kpc$ \citep{Zhuravleva2012}.

If gas motions can be effectively described by a single driver (i.e., a dominant injection scale $\ell_{\rm inj}$), the PDFs of the velocity ratio reflect the shape of the velocity power spectrum. We characterize this effect using Gaussian random velocity fields with an underlying energy power spectra \citep{Subramanian2006,Zhang2025}
\be
E(k) = E_0 \frac{(k\ell_{\rm inj})^{\alpha}}{1+(k\ell_{\rm inj})^{\alpha-2}}e^{-k\ell_{\rm diss}},
\label{eq:Ek}
\ee
where $E_0$ is a normalization factor, $\ell_{\rm diss}$ is the dissipation scale, and $\alpha$ is the spectral index in the inertial range. We explore a range of $E(k)$ forms and estimate their corresponding velocity-ratio PDFs assuming the Coma gas density radial profile \citep{Churazov2012}. The results are shown in Fig.~\ref{fig:turb_sim}, where the PDFs are computed within $250\kpc$ (approximately the Coma core size) and averaged over 1000 realizations. As expected, the PDFs are symmetric around zero. The distributions become broader when large-scale motions (larger than $\ell_{\rm eff}$) are present, which contribute more power to the bulk velocities. A PDF standard deviation of order unity generally indicates that $\ell_{\rm inj}>\ell_{\rm eff}$ and the cascade is well developed \citep{Zhuravleva2012}. This corresponds to the case of quasi-relaxed clusters, where mergers have driven gas motions on large scales in the past, such as the Perseus cluster and A2029 \citep{XRISM2025_A2029_Outer,Zhang2025}.

In the bottom panel of Fig.~\ref{fig:turb_sim}, we overlay the three XRISM measurements of this ratio in Coma \citep{XRISM2025_Coma,Gatuzz2025}, which are all negative and scattered over a broad range. They are statistically unlikely to be interpreted as arising from a Gaussian random field, even for large $\ell_{\rm inj}$ and a steep spectral slope. We note that the mean bulk velocity of the three XRISM measurements ($\simeq-500\kms$) provides critical information on gas motions, which is however automatically cancelled in estimates of the velocity structure function \citep{XRISM2025_Coma,Eckert2025}.

In systems where turbulence has had sufficient time to develop, it is unlikely that significant gas bulk motions (e.g., $|u_{\rm los}/\sigma_{\rm los}|\gg1$) arise on scales much larger than $\ell_{\rm eff}$ (approximately the pressure scale height), since gravitational stratification limits gas motions in the radial direction. In contrast, merging systems naturally produce velocity fields whose velocity-ratio PDFs deviate significantly from a symmetric Gaussian (see Fig.~\ref{fig:ratio_sim}). However, this phase is transient. Tailored numerical models of the target are essential for interpreting the measurements in such a phase. As shown in our simulation, the merger-driven velocity field eventually evolves into a developed state within $\simeq4\Gyr$.

\subsection{Absence of cool core in Coma and impact of previous mergers}

As evidenced by the presence of two very bright central galaxies and the absence of a cool core, the Coma cluster went through a significant merger in the past.
Numerical simulations suggested that nearly head-on mergers with small to moderate mass ratios (e.g., $\xi\lesssim10$) can transfer clusters from cool-core to non-cool-core systems \citep[e.g.,][]{ZuHone2011,Hahn2017}, and such mergers generally require $\sim5\Gyr$ to reach complete relaxation \citep[e.g.,][]{Poole2006,Zhang2016}. These results indicate that the off-axis minor merger discussed in this paper is unrelated to Coma's non-cool core formation. The mergers responsible for it likely occurred several Gyrs ago, as the major observed X-ray morphological features can be sufficiently explained by our minor merger. This argument is further supported by the cosmological simulation with constrained initial conditions SLOW \citep{Dolag2023}, which reproduces the observed large-scale structure of the local Universe. The Coma analogue in this simulation undergoes a rapid growth from $z\sim1.5$ and likely had its most recent significant merger at $z\sim0.25$ (Klaus Dolag, priv. comm.). Any strong velocity gradients, as those observed in the current Coma, are unlikely to persist if they had been generated several dynamical timescales earlier (see Fig.~\ref{fig:ratio_sim}). Besides, our simulations predict $\sigma_{\rm los}\simeq100-150\kms$ near the cluster core. Assuming that, to the first approximation, the line broadening $\sigma_0$ due to pre-merger turbulence has to be added quadratically, the observed broadening $\sim 200\kms$ suggests $\sigma_0\sim 130-170\kms$, albeit with large uncertainties. This result provides yet another hint that earlier mergers do not dominate the present-day velocity field.

\section{Conclusions}

In this letter, we demonstrate that the most salient features of the velocity field in the recent XRISM measurements of the Coma cluster core can be explained with an off-axis minor merger.

In our scenario, the infalling group (NGC~4839) bypasses Coma and induces large-scale bulk motions in the cluster core, which did not have sufficient time to fully cascade into turbulence. In contrast, the turbulence is already well developed in the group’s wake along its trajectory; however, this turbulent wake does not intersect the central region of Coma because of the non-zero impact parameter. This naturally explains why the velocity dispersion in the Coma core remains moderate ($\simeq200\kms$) despite the ongoing energetic merger.

We further demonstrate that the ratio between the LOS velocity relative to galaxies and the velocity dispersion provides a simple and robust diagnostic of the type and scale of gas motions. We show that the velocity ratios in Coma deviate significantly from Gaussian random field predictions, consistent with its unrelaxed dynamical state.

\begin{acknowledgements}

CZ thanks the host at the Max Planck Institute for Astrophysics for his visit and the inspiring discussions with Frederick Groth, Ulrich Steinwandel, and Klaus Dolag. CZ and NW were supported by the GACR EXPRO grant No. 21-13491X. CZ and IZ acknowledge partial support from the Alfred P. Sloan Foundation through the Sloan Research Fellowship. CZ and IZ were partially supported by NASA grant 80NSSC18K1684. IK acknowledges support by the COMPLEX project from the European Research Council (ERC) under the European Union’s Horizon 2020 research and innovation program grant agreement ERC-2019-AdG 882679. The simulations presented in this paper were carried out using the Midway computing cluster provided by the University of Chicago Research Computing Center.

\end{acknowledgements}


\begin{appendix} %First appendix
\onecolumn

\section{Simulation setups} \label{sec:appendix:methods}

\begin{figure*}[!h]
\centering
\includegraphics[width=0.95\linewidth]{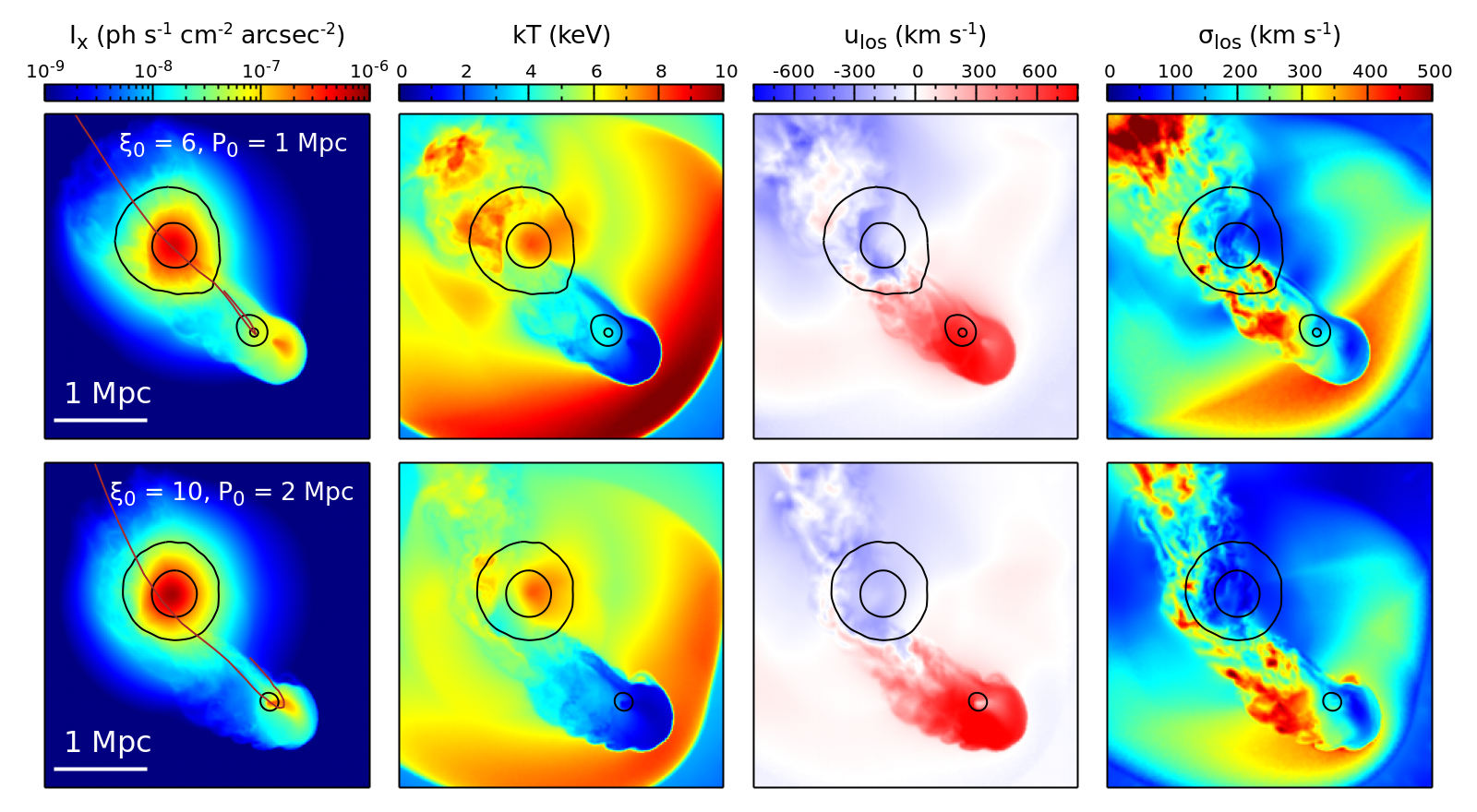}
\vspace{-5pt}
\caption{Same as Fig.~\ref{fig:merger_proj}, but for simulations with different merger mass ratios and initial impact parameters (both with $V_0=500\kms$). For simplicity, we adopt the same inclination angle for the projection as in Fig.~\ref{fig:merger_proj}. These examples illustrate that slightly varying the merger parameters produces broadly similar X-ray morphologies and velocity patterns.}
\label{fig:merger_proj_params}
\end{figure*}

We followed \citet{Lyskova2019} for the numerical setups in this work (see also \citealt{Zhang2014} for more details). In each simulation, we model mergers between two idealized cool-core clusters. Each cluster consists of gas and dark matter components that are spherical and in equilibrium in their initial conditions. The mass of the main cluster is fixed at $M_\mathrm{vir} = 1.2\times10^{15}\,M_\odot$ in all simulations, motivated by weak-lensing measurements \citep{Okabe2010}. In our baseline model (see Fig.~\ref{fig:merger_proj}), we adopt a slightly smaller merger mass ratio ($\xi_0=6$) than the value used in \citet[][$\xi_0=10$]{Lyskova2019}, which reproduces a more prominent bow shock ahead of the infalling group. This choice, however, does not affect any of the conclusions in this paper. The other merger parameters, namely the initial pairwise velocity ($V_0=500\kms$) and impact parameter ($P_0=2\Mpc$), are kept the same as suggested in \citet{Lyskova2019}. We additionally include a passively advected scalar field to trace gas that was initially associated with the infalling group. Fig.~\ref{fig:merger_proj_params} presents simulations with varied merger parameters. They both show broadly similar X-ray morphologies and velocity patterns as in Fig.~\ref{fig:merger_proj}, demonstrating that our results are robust and do not require extreme fine-tuning of merger parameters to match the observations.

\section{Gas motions in the merger plane} \label{sec:appendix:sloshing}

\begin{figure}
\centering
\includegraphics[width=0.6\linewidth]{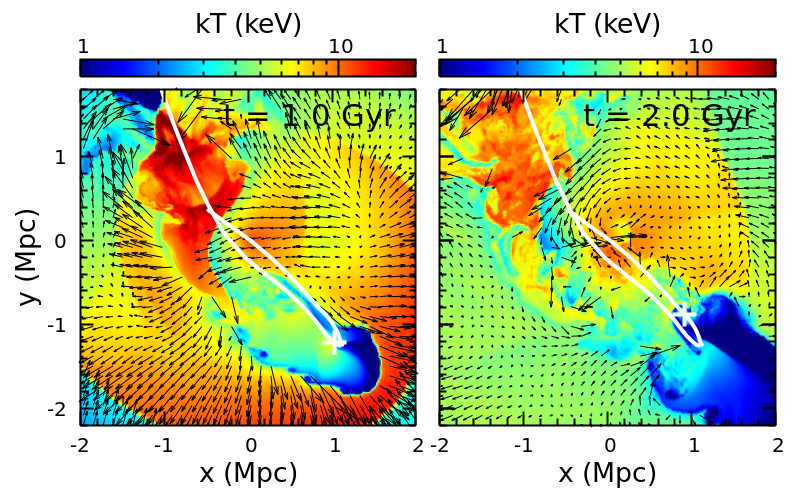}
\vspace{-5pt}
\caption{Temperature slices across the merger plane at $t=1\Gyr$ (left) and $t=2\Gyr$ (right). The overlaid vectors indicate the amplitude and direction of the gas velocity field (in the merger plane) within the atmosphere dominated by the main cluster. The white solid lines show the trajectory of the infalling group, with white crosses marking its current position. This figure demonstrates that the bypassing group drives large-scale bulk motions in the core of the Coma cluster.}
\label{fig:slice}
\end{figure}

Fig.~\ref{fig:slice} shows slices of the gas temperature across the merger plane at $t=1$ and $2\Gyr$ from the simulation shown in Fig.~\ref{fig:merger_proj}, overlaid with gas velocity vectors that illustrate both the direction and amplitude of the in-plane motions. In the central region, most velocity vectors point to the right, explaining the moderate velocity dispersion. Over a timescale of $\sim1\Gyr$, these bulk motions slowly evolve into large-scale rotations as a result of angular momentum injected by the group, eventually producing gas sloshing (see \citealt{Zuhone2016} for a review). At the current epoch ($t\simeq1\Gyr$), high velocity dispersion ($\gtrsim300\kms$) appears in the group's wake/stripped tail, where turbulence is well developed. Meanwhile, the high velocity dispersion downstream of the runaway merger shock is largely produced by gas expansion.

\end{appendix}

\end{document}